\documentclass{article}
\usepackage{arxiv}

\usepackage[utf8]{inputenc}
\usepackage[T1]{fontenc}
\usepackage{hyperref}
\usepackage{graphicx}
\usepackage{amsmath, amsfonts}

\usepackage[numbers]{natbib}

\usepackage{doi}
\usepackage{booktabs}
\usepackage{caption}

\title{Retrieval-Augmented Generation Assistant for Anatomical Pathology Laboratories}

\author{
\begin{tabular}{ccc}
    \href{https://orcid.org/0009-0001-2611-4239}{\includegraphics[scale=0.06]{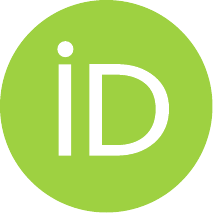}\hspace{1mm}Diogo Pires} &
    \href{https://orcid.org/0009-0004-1046-7883}{\includegraphics[scale=0.06]{orcid.pdf}\hspace{1mm}Yuriy Perezhohin} &
    \href{https://orcid.org/0000-0002-8793-1451}{\includegraphics[scale=0.06]{orcid.pdf}\hspace{1mm}Mauro Castelli} \\
    {\small ORCID: \href{https://orcid.org/0009-0001-2611-4239}{0009-0001-2611-4239}} &
    {\small ORCID: \href{https://orcid.org/0009-0004-1046-7883}{0009-0004-1046-7883}} &
    {\small ORCID: \href{https://orcid.org/0000-0002-8793-1451}{0000-0002-8793-1451}} \\
    & {\small Remynd, Alameda Bonifácio Lazaro Lozano, nº15, 1ºC} & \\
    & {\small Oeiras, Portugal} & \\
\end{tabular}
\\[2.0em]
\parbox{0.85\textwidth}{%
  \centering
  \textit{Nova Information Management School (NOVA IMS), Universidade Nova de Lisboa,}\\
  \textit{Campus de Campolide, 1070-312 Lisboa, Portugal}%
}
}

\date{}

\begin{document}

\maketitle

\begin{abstract}
Accurate and efficient access to laboratory protocols is essential in Anatomical Pathology (AP), where up to 70\% of medical decisions depend on laboratory diagnoses. However, static documentation such as printed manuals or PDFs is often outdated, fragmented, and difficult to search, creating risks of workflow errors and diagnostic delays. This study proposes and evaluates a Retrieval-Augmented Generation (RAG) assistant tailored to AP laboratories, designed to provide technicians with context-grounded answers to protocol-related queries. 
We curated a novel corpus of 99 AP protocols from a Portuguese healthcare institution and constructed 323 question-answer pairs for systematic evaluation. Ten experiments were conducted, varying chunking strategies, retrieval methods, and embedding models. Performance was assessed using the RAGAS framework (faithfulness, answer relevance, context recall) alongside top-$k$ retrieval metrics.
Results show that recursive chunking and hybrid retrieval delivered the strongest baseline performance. Incorporating a biomedical-specific embedding model (MedEmbed) further improved answer relevance (0.74), faithfulness (0.70), and context recall (0.77), showing the importance of domain-specialised embeddings. Top-$k$ analysis revealed that retrieving a single top-ranked chunk ($k=1$) maximized efficiency and accuracy, reflecting the modular structure of AP protocols.
These findings highlight critical design considerations for deploying RAG systems in healthcare and demonstrate their potential to transform static documentation into dynamic, reliable knowledge assistants, thus improving laboratory workflow efficiency and supporting patient safety.
\end{abstract}


\section{Introduction}
Healthcare technicians in Anatomical Pathology (AP) laboratories must navigate an extensive and constantly evolving set of biomedical protocols and equipment manuals. Given that approximately 70\% of medical decisions depend on laboratory diagnoses \cite{WHO2019}, ensuring accuracy and efficiency in AP workflows is essential. Detailed procedural methodologies are critical for guiding technicians and pathologists through key biomedical tasks, including tissue processing, sample staining, and equipment maintenance \cite{Bancroft2019, Paulino2022}. Yet, keeping these protocols both up-to-date and easily accessible presents a persistent challenge. The rapid introduction of new technologies and instruments demands continuous learning and frequent manual consultation, often under time pressure. Traditional resources such as printed manuals or static PDFs, while still widely used, are cumbersome to navigate and quickly become outdated \cite{Labware2021, eLabNext2024}. As a result, laboratory staff may face difficulties in retrieving precise procedural information when it is most needed, thereby increasing the risk of workflow disruptions and diagnostic errors.

Current applications of Artificial Intelligence (AI) assistants in healthcare have largely focused on patient-facing services, such as chatbots for handling medical inquiries, or on administrative tasks, including scheduling and documentation automation \cite{Dammavalam2022, Yang2023}. While valuable, these uses overlook a critical opportunity: supporting healthcare professionals directly in their daily practice. In AP laboratories, for example, technicians may urgently need to prepare a rarely used chemical reagent for a highly specific staining procedure. In such cases, any delay or failure to locate the correct protocol risks producing inaccurate results, with potentially serious consequences for patient care. An AI assistant tailored for laboratory environments could bridge this gap by providing immediate, context-specific guidance, delivering procedural steps on demand, and answering questions related to techniques, protocols, and equipment use.

Recent developments in healthcare AI have demonstrated the potential of RAG systems \cite{oche2025, jang2025} across various clinical domains. Studies have shown RAG's effectiveness in clinical documentation assistance, medical question answering, and diagnostic support \cite{gargari2025, amugongo2025}. For instance, Quidwai \& Lagana \cite{quidwai2024} developed a RAG system for retrieving treatment guidelines from oncology databases, while Bernardi \& Cimitile \cite{bernardi2024} demonstrated improved accuracy in radiology report generation through context-aware retrieval. However, these applications have predominantly focused on physician-facing tools or patient communication systems.

A critical gap remains in supporting laboratory personnel who execute the technical procedures underlying diagnostic workflows. In anatomical pathology specifically, where protocol adherence directly impacts diagnostic quality, no prior work has evaluated RAG systems for real-time procedural guidance. Unlike clinical decision support, which often requires synthesizing evidence across multiple sources, laboratory protocol retrieval demands precise, single-source answers to specific technical questions. This distinction requires different design considerations in chunking, retrieval, and evaluation strategies, aspects that have not been systematically explored in previous RAG implementations.

This paper addresses this gap by developing a Retrieval-Augmented Generation (RAG) assistant specifically designed for AP laboratories. The system builds on recent advances in Large Language Models (LLMs) to deliver precise, context-aware answers to technicians' procedural queries. To support this effort, we introduce a novel corpus of AP laboratory protocols, curated from a Portuguese healthcare institution, together with a benchmark set of Question-Answer (QA) pairs for systematic evaluation. Using this dataset as the retrieval backbone, we conducted a series of experiments exploring variations in document chunking strategies, retrieval methods, and embedding models to identify the most effective pipeline configuration for this domain. Our contributions are threefold: (1) The creation and public release of a new AP Protocols dataset for the research community; (2) An empirical analysis of how retrieval design choices affect quality, relevance, and faithfulness in generated answers; (3) A demonstration of how domain-specific RAG systems can directly enhance laboratory workflows by transforming static protocols into dynamic, queryable knowledge resources.

The remainder of this paper is organized as follows. Section 2 reviews related work across three domains: AP laboratory practices, large language models in biomedicine, and retrieval-augmented generation systems. Section 3 describes our methodology, including corpus construction, chunking strategies, retrieval mechanisms, embedding models, experimental design, and evaluation metrics. Section 4 presents and discusses experimental results, analyzing the impact of different design choices on the performance of the system. Section 5 concludes with a summary of findings, acknowledges limitations, and proposes directions for future research.

\section{Related Work}
Research relevant to this study spans three domains: the organisation and standardisation of AP laboratory protocols, the development of LLMs for biomedical applications, and the use of RAG frameworks to improve accuracy and faithfulness. While each area has advanced significantly, their integration to support AP workflows remains largely unexplored.

\subsection{Anatomical Pathology}
Anatomical Pathology laboratories, including histology and cytology services, are central to diagnostic medicine, where tissue, cell, and fluid samples are examined to detect neoplasia and other diseases \cite{Bancroft2019}. The accuracy of AP results directly influences clinical decision-making and patient outcomes, making standardised and reproducible protocols indispensable for maintaining quality \cite{Paulino2022}. Technicians are required to master a wide array of procedures, ranging from tissue fixation, embedding, and sectioning to staining, microscopy screening, and equipment calibration. Rotations across laboratory sections further underscore the need for consistent access to comprehensive, up-to-date procedural knowledge.

However, maintaining such documentation is a persistent challenge. The introduction of new equipment often necessitates rapid assimilation of revised operating and maintenance instructions. Historically, laboratories relied on printed manuals, which suffered from version-control problems, physical deterioration, and poor searchability \cite{Labware2021}. The transition to digital documents and static PDFs has alleviated some of these issues but continues to impose inefficiencies: protocols can still be outdated, scattered across systems, and time-consuming to navigate \cite{eLabNext2024}. These limitations raise the risk that incorrect or obsolete procedures may be applied, potentially compromising diagnostic accuracy. Consequently, there is increasing interest in intelligent, dynamic knowledge systems capable of retrieving the most relevant and current information on demand \cite{Pillay2025}. In this context, an AI assistant tailored to AP workflows could function as a real-time knowledge aide, providing instant access to procedural guidance, reducing cognitive burden on staff, and lowering the likelihood of human error.

\subsection{Large Language Models}
Large Language Models are advanced AI systems that have transformed Natural Language Processing (NLP) by enabling high-performance text summarisation, translation, question answering, and conversational agents \cite{Jurafsky2024}. Built on the Transformer architecture \cite{Vaswani2017}, LLMs rely on encoder-decoder structures and self-attention mechanisms to capture long-range dependencies and semantic relationships in text. Pre-training on massive corpora equips these models with broad linguistic competence, which can then be specialised through fine-tuning on domain-specific datasets \cite{Raffel2023}. In the biomedical domain, this approach has led to models such as BioBERT, which improved downstream performance by incorporating specialised medical texts \cite{Lee2020}.

Applications of LLMs in healthcare are diverse, ranging from generating patient notes and summarizing electronic health records to supporting diagnostic reasoning \cite{Park2024, Wang2024}. However, their direct deployment in clinical and laboratory environments poses significant risks \cite{Sudhi2024}. LLMs are prone to hallucinations (plausible but incorrect outputs), especially when the necessary domain-specific knowledge is absent from their training corpus \cite{Yang2023, Niu2024}. This limitation is particularly problematic in high-stakes biomedical contexts, where factual inaccuracies can have direct consequences for patient care.

To mitigate these risks, three complementary strategies have been explored: (1) Prompt engineering, which refines input queries to guide LLMs toward more accurate and structured outputs \cite{Marvin2024}; (2) Fine-tuning, which adapts general-purpose models to domain-specific tasks using targeted datasets \cite{Howard2018}; (3) Knowledge retrieval, which augments models with access to external information sources, ensuring that responses are grounded in verified knowledge. In this category, Retrieval-Augmented Generation has emerged as one of the most promising approaches for ensuring both accuracy and faithfulness in sensitive domains \cite{Lewis2021}. In particular, this method grounds the LLM’s responses in up-to-date domain knowledge, reducing hallucinations and improving accuracy.

\subsection{Retrieval-Augmented Generation}
\begin{figure}[!h]
    \centering
    \includegraphics[width=0.95\linewidth]{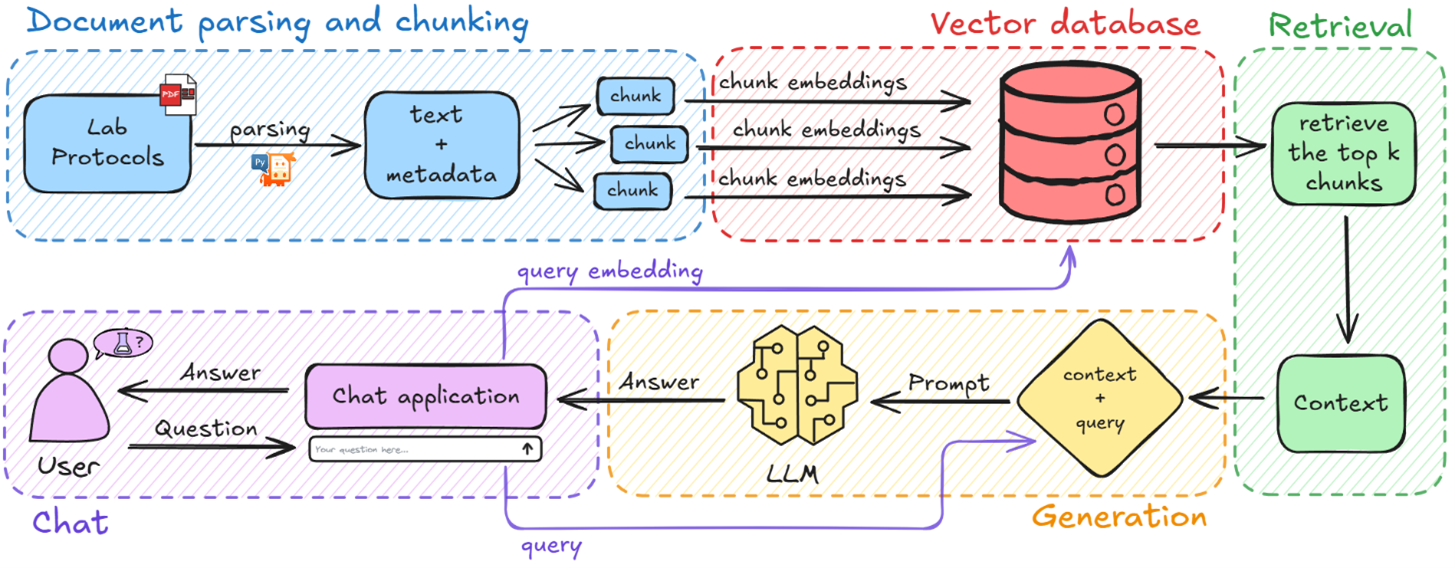}
    \captionsetup{justification=centering}
    \caption{Diagram of a typical RAG pipeline.}
    \label{fig:rag_pipeline}
\end{figure}

Retrieval-Augmented Generation (RAG) integrates information retrieval with generative modelling to produce answers that are both contextually grounded and dynamically updated. In a typical RAG pipeline (Figure \ref{fig:rag_pipeline}), domain-specific documents are divided into smaller chunks, embedded into a vector space, and stored in a vector database. User queries are embedded in the same space, and the most relevant chunks, determined via similarity metrics such as cosine similarity, are retrieved and appended to the LLM prompt. The model then generates a response that is explicitly conditioned on the retrieved evidence.

This framework directly addresses two major limitations of standalone LLMs: the knowledge cutoff inherent to pre-training and the tendency to generate factually inaccurate responses \cite{Lewis2021}. By grounding outputs in up-to-date external knowledge, RAG systems provide both factual accuracy and traceability, qualities essential for biomedical applications.

Recent studies have demonstrated the value of RAG in healthcare. For instance, MedRAG was designed to assist clinicians by retrieving diagnosis and treatment recommendations from curated medical knowledge graphs, thereby improving interpretability and reasoning \cite{medrag2025}. These efforts illustrate the potential of RAG-based assistants to align generated answers with authoritative biomedical sources.

Our study extends this line of work by applying RAG to the largely unexplored domain of AP laboratory protocols. By constructing a retriever over a comprehensive corpus of AP documents, our system allows technicians to query procedures and receive responses directly linked to official lab documentation. This design not only improves procedural efficiency but also ensures that every recommendation is anchored in verifiable sources, thus building trust among laboratory professionals and supporting high standards of diagnostic reliability.

\section{Methodology}
This section describes the methodological framework adopted to develop and evaluate the proposed RAG assistant for AP laboratories. We detail the construction of the document corpus, preprocessing and chunking strategies, retrieval mechanisms, embedding models, and the experimental setup. Finally, we present the evaluation metrics used to assess both retrieval accuracy and the faithfulness of generated responses. Together, these steps establish a reproducible pipeline for testing the impact of different design choices on RAG performance in a biomedical laboratory setting.

At a high level, the RAG assistant is organised as a modular pipeline: a curated and anonymised AP corpus is segmented into coherent chunks, embedded, and indexed in \textit{ChromaDB}, while user queries retrieve relevant context via dense, reranked, and hybrid search. A generation layer then produces answers grounded in the retrieved passages, with chunking, embedding, and retrieval modules remaining interchangeable to support easy adaptation to laboratory needs.

\subsection{Document Corpus}
We constructed a corpus of 99 Anatomical Pathology documents obtained from a Portuguese healthcare institution. 

The corpus covered routine procedures, specimen processing techniques, staining protocols, equipment operation, and maintenance instructions. Full text and metadata, including protocol category, title, and keywords, were extracted using \textit{PyMuPDF}. Preprocessing steps removed redundant elements such as headers, footers, and confidential identifiers. Sensitive or identifying information was anonymised to comply with privacy and data-protection standards.

For reproducibility and community use, the final dataset was formatted into a HuggingFace Datasets compatible structure, where each entry consists of the document's Portuguese text and associated metadata. This design facilitates future extensions, such as fine-tuning domain-specific models or conducting retrieval-based benchmarks.

\subsection{Chunking}
Feeding entire documents into the RAG pipeline is inefficient and risks diluting retrieval precision. Instead, we implemented chunking strategies to segment documents into smaller, semantically meaningful units of text \cite{Qu2024}. Two approaches were explored: recursive chunking and semantic chunking.

Recursive chunking uses a length-based splitting method with configurable overlaps. We employed LangChain's \textit{RecursiveCharacterTextSplitter} with target sizes of $\sim$256 and $\sim$512 tokens and overlaps of 64 and 128 tokens, respectively. By specifying natural breakpoints such as newlines and punctuation, this approach minimized mid-sentence splits.

Semantic chunking aims to preserve contextual integrity by splitting text according to semantic boundaries, rather than fixed length. Implemented via LangChain's \textit{SemanticChunker}, it applies an embedding-based anomaly detection method to identify semantic discontinuities \cite{langchain_semantic2025}. We used a minimum chunk size of $\sim$128 tokens and a default semantic shift threshold of 95. This method is particularly suitable for AP protocols, where maintaining context (e.g., within staining procedures or equipment instructions) is essential \cite{langchain_semantic2025}.

\subsection{Retrieval}
Retrieval constitutes the core of the RAG pipeline. Document chunks were embedded into vector representations and indexed within ChromaDB, with separate databases for each chunking strategy. Queries were embedded using the same model, and relevant chunks were retrieved through three strategies: naïve retrieval, reranking retrieval, and hybrid search.
Naïve retrieval served as the baseline, ranking chunks by cosine similarity between query and chunk embeddings. Specifically, it uses cosine similarity between the query vector q and each document chunk vector $c_i$:
\begin{equation}
    \cos(\theta)=\frac{(q\cdot c_i)}{||q||\space||c_i||}
\end{equation}

This score indicates how closely the chunk's content matches the query in semantic terms, where $\theta$ represents the angle between the query vector q and the chunk vector $c_i$ in the embedding space, $q \cdot c_i$ represents their dot product, and $||q||$ and $||c_i||$ represent the Euclidean norms (magnitudes) of the query and chunk vectors, respectively. The system then retrieves the top-$k$ chunks with the highest cosine similarity scores. In our experiments, we set $k=3$, meaning the top three chunks are used as the context for answer generation.

To improve retrieval precision over the baseline, we adopted a second strategy by implementing a reranking step designed to filter out less relevant contexts. With reranking retrieval, we initially retrieve a set of candidate chunks (similar to the naïve approach) and then introduce an additional filtering criterion based on a similarity score threshold. Specifically, by establishing a threshold of 0.40, we select only those chunks whose cosine similarity scores relative to the user's query surpass this threshold.

The hybrid search retrieval strategy combines dense semantic search with sparse keyword search to leverage their complementary strengths. In our implementation, we augment the embedding-based results with a lexical search using the BM25 algorithm for term matching. The following equation shows the BM25 scoring function (Retrieval Status Value, $rsv$) for a given query $q$ and document $d$ \cite{Trotman2014}:

\begin{equation}
\label{hybrid_search_equation}
    rsv_q=\sum_{t\in q}\log \left(\frac{N}{df_t}\right) \cdot \frac{(k_1+1) \cdot tf_{td}}{k_1\cdot \left(1-b+b\cdot (\frac{L_d}{L_{avg}})\right)+tf_{td}}
\end{equation}

The summation is performed over all query terms $t$. Here, $N$ denotes the total number of documents (chunks) in the corpus, and $df_t$ is the document frequency (number of documents containing term $t$), $tf_{td}$  represents the term frequency of $t$ in document $d$, while $L_d$ and $L_{avg}$ correspond to the length of document $d$ and the average document length across the corpus, respectively. $k_1$ and $b$ are hyperparameters controlling term frequency saturation and document length normalization, respectively. In our implementation, we used the default values $k_1 = 1.5$ and $b = 0.75$, which are standard in BM25 applications.

Intuitively, BM25 assigns higher scores to documents that contain frequent occurrences of the query terms, while penalizing terms that appear in many documents (since they carry less discriminative power). At the same time, it corrects for length bias by avoiding over-rewarding long documents where query terms may appear often simply due to size. In this way, BM25 balances term importance, frequency saturation, and length normalisation to compute a robust relevance score \cite{Trotman2014}.

In practice, our implementation computes cosine similarity and BM25 scores and then combines the two methods using LangChain's \textit{EnsembleRetriever}. We configured this retriever by assigning 30\% weight to semantic relevance (dense search) and 70\% weight to keyword relevance (sparse search). Chunks that are top-ranked in either semantic similarity or BM25 score are considered for inclusion in the final top-$k$ results.

\subsection{Embedding Models}
The embedding model used to vectorise the text chunks and user queries is a critical component of the retrieval and evaluation pipeline. We evaluated two different embedding models in our pipeline: a general-purpose multilingual model and a biomedical-specific model.

For most experiments, we used the \textit{paraphrase-multilingual-MiniLM-L12-v}2 model from SentenceTransformers. This general-purpose multilingual model contains $\sim$118M parameters, mapping sentences into a 384-dimensional dense vector space optimised for semantic search \cite{Reimers2019}. This model was chosen due to its compact size, efficiency, and strong performance across multiple languages, providing a robust baseline for testing different chunking and retrieval methods on documents written in Portuguese.

\textit{MedEmbed-small-v0.1} from HuggingFace was used to evaluate potential performance gains when compared with the general-purpose embeddings. This domain-specific model is designed for medical and clinical NLP tasks, being able to capture the nuances and complexities of medical terminology and concepts. This makes the model particularly useful for biomedical information retrieval. Similar to the previous model, although significantly smaller ($\sim$33M parameters), MedEmbed maps sentences into a 384-dimensional dense embedding space; however, it generates embeddings specifically grounded in clinical language, thus potentially improving semantic matching for specialised biomedical content, as is the case with AP lab protocols \cite{MedEmbed2024}.

\subsection{Experimental Setup}
We conducted ten experiments exploring different combinations of chunking, retrieval, and embedding models. Experiments 1-3 tested naïve retrieval across the three chunking strategies. Experiments 4-6 applied reranking retrieval; Experiments 7-9 used hybrid search. Finally, Experiment 10 evaluated the biomedical-specific embedding model with the best-performing configuration.

All experiments were conducted on a local machine (AMD Ryzen 7 CPU, 16 GB RAM, NVIDIA GeForce RTX 3070 GPU). The RAG pipeline was implemented in Python using LangChain and the HuggingFace Transformers library. The model used for answer generation was Llama 3.1 8B (see https://ollama.com/library/llama3.1 documentation), a multilingual LLM from Meta with a 128K context length. This model was run locally using Ollama, allowing free and offline usage. Table \ref{tab:experiments} summarises all experimental configurations.

\begin{table}[h!]
    \centering
    \caption{Overview of Experimental Configurations}
    \label{tab:experiments}
    \begin{tabular}{cllll}
        \toprule
        \textbf{Exp.} & \textbf{Chunking} & \textbf{Retrieval} & \textbf{Embedding model} & \textbf{LLM} \\
        \midrule
        1  & Recursive 256 & Naïve         & \textit{paraphrase-multilingual} & Llama 3.1 \\
        2  & Recursive 512 & Naïve         & \textit{paraphrase-multilingual} & Llama 3.1 \\
        3  & Semantic      & Naïve         & \textit{paraphrase-multilingual} & Llama 3.1 \\
        4  & Recursive 256 & Reranking     & \textit{paraphrase-multilingual} & Llama 3.1 \\
        5  & Recursive 512 & Reranking     & \textit{paraphrase-multilingual} & Llama 3.1 \\
        6  & Semantic      & Reranking     & \textit{paraphrase-multilingual} & Llama 3.1 \\
        7  & Recursive 256 & Hybrid search & \textit{paraphrase-multilingual} & Llama 3.1 \\
        8  & Recursive 512 & Hybrid search & \textit{paraphrase-multilingual} & Llama 3.1 \\
        9  & Semantic      & Hybrid search & \textit{paraphrase-multilingual} & Llama 3.1 \\
        10 & Recursive 512 & Hybrid search & \textit{MedEmbed-small-v0.1}     & Llama 3.1 \\
        \bottomrule
    \end{tabular}
\end{table}

\subsection{Evaluation Metrics}
To quantitatively evaluate retrieval and generation performance, we created a structured test dataset consisting of 323 QA pairs directly derived from the AP lab protocols. Each pair contains a user question, its corresponding ground-truth answer, and an associated reference context from the original documents. Each question was processed by the RAG pipeline using each experimental configuration to retrieve context chunks and generate responses, resulting in an evaluation dataset.

Performance was systematically assessed using the Retrieval-Augmented Generation Assessment (RAGAS) framework and standard top-$k$ evaluation metrics. The RAGAS framework, specifically tailored for RAG systems, evaluates the relationship between retrieved contexts and generated answers using the LLM-as-a-judge strategy through three key metrics \cite{Es2023, ragas2025}:

\begin{itemize}
    \item Faithfulness: Is the response grounded in the retrieved context? Faithfulness, $F$, measures if the generated answers are supported by the retrieved contexts, mitigating hallucinations:
    \begin{equation}
        F = \frac{\textit{Number of supported statements}}{\textit{Total number of statements}}
    \label{eq:faithfullness}
    \end{equation}

    \item Answer Relevance: Does the response address the question? Answer Relevance, $AR$, assesses the degree to which the generated answer directly addresses the user's question by generating n potential questions $q_i$ based on the generated answer $a_s(q)$ and computing their semantic similarity, $sim(q,q_i)$, to the original question $q$:
    \begin{equation}
        AR = \left( \frac{1}{n} \right) *\sum_{i=1}^{n} \text{$sim$}(q, q_i)
    \label{eq:answer_relevance}
    \end{equation}

    \item Context Recall: Is the retrieved context sufficiently focused? Context Recall, $CR$, evaluates the retrieval system’s ability to capture all necessary context chunks for accurate answer generation:
    \begin{equation}
        CR=\frac{\textit{Relevant retrieved chunks}}{\textit{Total number of retrieved chunks}}
    \label{eq:context_relevance}
    \end{equation}
    
\end{itemize}

In addition to evaluating retrieval and answer correctness using the LLM-as-a-judge framework, we independently assessed retrieval performance with a deterministic approach: top-$k$ evaluation. This method quantifies how well the retrieved contexts align with the ground-truth reference contexts extracted from the original documents. Specifically, it employs \textit{Precision@$k$}, \textit{Recall@$k$}, and \textit{F1-Score@$k$} to measure retrieval accuracy, recall coverage, and the balance between the two, without relying on semantic similarity judgments by LLMs.


We computed these metrics for multiple cutoff values ($k = 1, 2, 4, 8$) to examine how retrieval effectiveness evolves as the number of retrieved chunks increases. These values were selected to reflect practical constraints in RAG pipelines, where only a limited number of chunks can realistically be passed to the LLM without overloading its context window or diluting relevance. This multi-scale analysis provides insight into the trade-off between precision and recall at different retrieval depths. Taken together, these measures offer a comprehensive evaluation of both the retrieval module and the overall generative performance of the proposed RAG system.
 
\section{Results and Discussion}
This section presents the results of the experimental campaign. The experiments systematically evaluated the impact of chunking strategies, retrieval methods, embedding models, and top-$k$ configurations on RAG performance for AP protocols.

\subsection{Chunking: Recursive vs. Semantic}
Analysing the results of the chunking experiments with the general-purpose embedding model (Figure \ref{fig:chunking-retrieval}, left panel) indicates that, on average, the recursive methods outperformed semantic chunking across all metrics. Both recursive chunking strategies tested – 256 and 512-token chunks – achieved higher answer relevance (0.69 and 0.66), faithfulness (0.65 and 0.64), and context recall (0.58 and 0.70) compared to semantic chunking (which scored 0.57, 0.36, and 0.29 on these metrics, respectively). Although the 256-token chunks slightly edged the 512-token chunks in answer relevance (0.69 vs. 0.66) and faithfulness (0.65 vs. 0.64), the larger chunks offered a significantly higher context recall (0.58 vs. 0.70) by capturing more relevant text in each retrieved context.

The semantic chunking approach struggled noticeably, underperforming on all key metrics. We can attribute its poor performance to the highly uneven and excessive length of the chunks, producing some with over 2,800 tokens. Such oversized chunks dilute the concentration of relevant information and likely hinder the retrieval process, as the model must sift through too much text to find the answer. For example, consider a technician who needs a specific step from a tissue staining protocol. A semantic chunking strategy, as configured in our experiments, might return an overly large text chunk (e.g., an entire protocol of the staining manual) that includes the step but also many unrelated details. The RAG assistant’s answer would then have to navigate superfluous information, increasing the risk of confusion or missing the crucial detail. In contrast, with recursive chunking, the system would likely retrieve a concise snippet focused on that particular step, enabling a more direct and faithful answer. We can also conclude that the embedding model used to detect semantic boundaries lacked the capacity to identify optimal split points. A more advanced or fine-tuned model might have produced more coherent semantic chunks. 

To illustrate this limitation concretely, consider a typical use case: a technician preparing a Hematoxylin and Eosin (H\&E) staining protocol who needs to know the precise incubation time for the eosin step. With recursive 512-token chunking, the system would likely retrieve a focused chunk containing only the staining procedure steps, producing a response such as “Incubate slides in eosin solution for 2-3 minutes at room temperature”. However, with semantic chunking producing chunks exceeding 2,800 tokens, the retrieved context might include the entire staining manual, including reagent preparation, equipment setup, safety procedures, and quality control measures. This dilution forces the LLM to parse through extensive irrelevant information, increasing the risk of extracting incorrect details. In a laboratory setting where precision is the main concern, such errors could compromise specimen quality or diagnostic accuracy. Furthermore, the computational overhead of processing oversized chunks impacts system responsiveness, a critical factor when technicians need immediate guidance during active procedures. 

Overall, our results indicate that a simple recursive chunking strategy with a moderate chunk size ($\sim$512 tokens) provides the best balance between preserving context and maintaining answer relevance and faithfulness. In contrast, the semantic chunking method, as configured, undermined retrieval quality on all key metrics.

\subsection{Retrieval: Naïve vs. Reranking vs. Hybrid Search}
\begin{figure}[!h]
    \centering
    \includegraphics[width=0.99\linewidth]{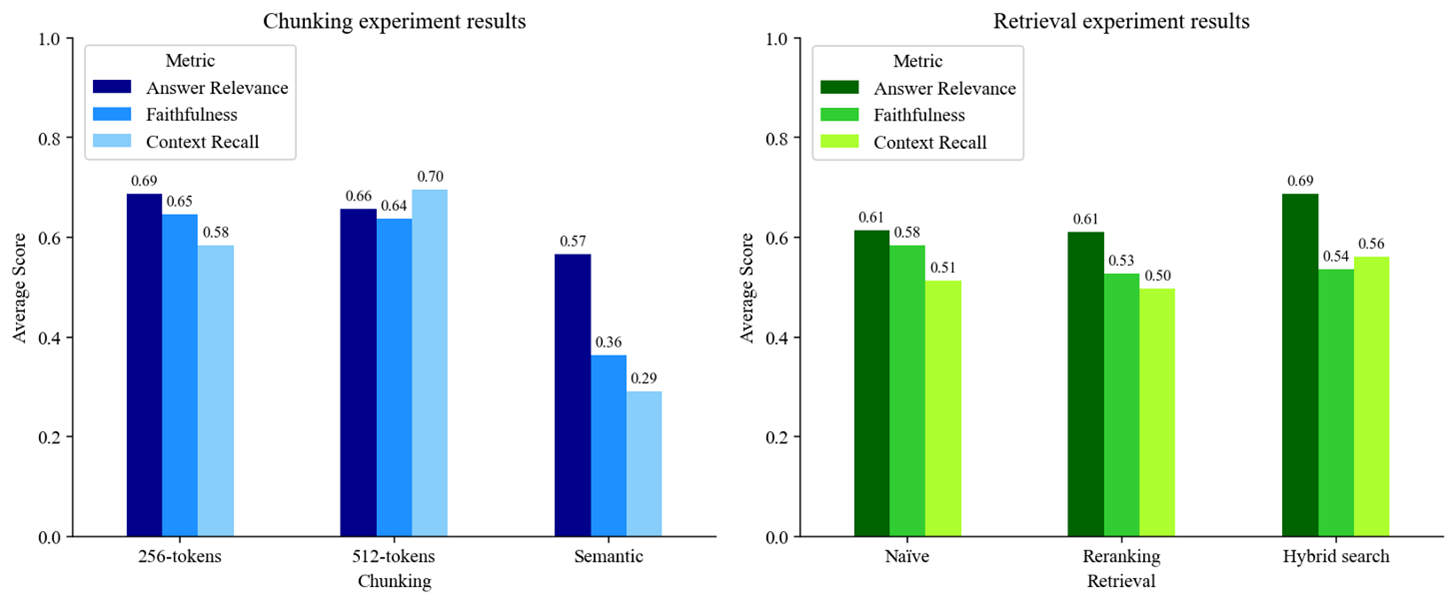}
    \captionsetup{justification=centering}
    \caption{Results of Chunking (left) and Retrieval (right) experiments, averaged across Experiments 1–9.}
    \label{fig:chunking-retrieval}
\end{figure}

The results (Figure \ref{fig:chunking-retrieval}, right panel) of the retrieval experiments show that, on average, the hybrid search strategy performed best overall, achieving the highest answer relevance (0.69) and context recall (0.56) among the three methods. The baseline naïve retrieval scored an answer relevance of 0.61 and context recall of 0.51, while the reranking strategy reached similar values (0.61 and 0.50, respectively). Notably, the naïve method did slightly outperform hybrid search in one aspect: answer faithfulness (0.59 vs. 0.54), indicating that the straightforward semantic-only retrieval sometimes yielded answers more grounded in the provided context.

The reranking method did not meet expectations. It provided no improvement in answer relevance (0.61, identical to naïve) and actually underperformed the baseline in context recall (0.50 vs. 0.51) and faithfulness (0.53 vs. 0.58). This underwhelming performance can be attributed to its filtering of potentially useful context. In our implementation, applying a fixed similarity threshold (cosine similarity score > 0.40) likely pruned out some chunks that contained relevant information just below the cutoff. This meant that the model was sometimes left with too little context to fully answer the user's question.

Overall, adding the reranking layer of complexity did not yield benefits in our setting, whereas the hybrid method (combining semantic understanding with keyword matching) retrieved more pertinent context without any meaningful loss of grounding. In practice, hybrid search is advantageous when terminology varies between the user’s query and the protocol text. For example, a technician may ask about “microtome calibration”, while the maintenance document uses the phrase “microtome adjustment”. A purely semantic (naïve) retriever may not score this context as high, and a reranking filter could even drop it if its similarity score falls just below the threshold. The hybrid approach, however, leverages keyword matching (with BM25) to anchor on “microtome” and related terms while dense retrieval supplies semantic proximity, ensuring the correct maintenance chunk is retrieved. This combination reliably retrieves the precise procedural steps needed, improving completeness and practical utility for AP workflows.

The superior context recall of hybrid search becomes particularly valuable in scenarios involving terminology variability. AP protocols often use equipment-specific terminology (e.g., “Leica ASP300S automated tissue processor”) alongside generic descriptors (“automated processor”). When a technician queries using generic terms, pure semantic retrieval may assign lower similarity scores to equipment-specific chunks, potentially missing the most relevant maintenance protocol. Hybrid search mitigates this by combining BM25's keyword matching, which captures exact equipment names, with semantic understanding of functional equivalence. This dual approach ensures retrieval of both precisely matching and semantically related content. However, the slightly reduced faithfulness score (0.54 vs. 0.59 for naïve retrieval) warrants consideration. Analysis of individual responses revealed that hybrid search occasionally retrieved multiple chunks with overlapping but not identical information (e.g., different versions of a protocol or related procedures), leading the LLM to synthesize information across sources. While this increased answer completeness, it sometimes introduced minor inconsistencies not present in single-source responses. For laboratory applications where strict adherence to official protocols is essential, this tradeoff suggests potential value in implementing source verification mechanisms or prioritizing chunks from the most recently updated documents.

\subsection{Embeddings: General vs. Biomedical}
\begin{figure}[!h]
    \centering
    \includegraphics[width=0.80\linewidth]{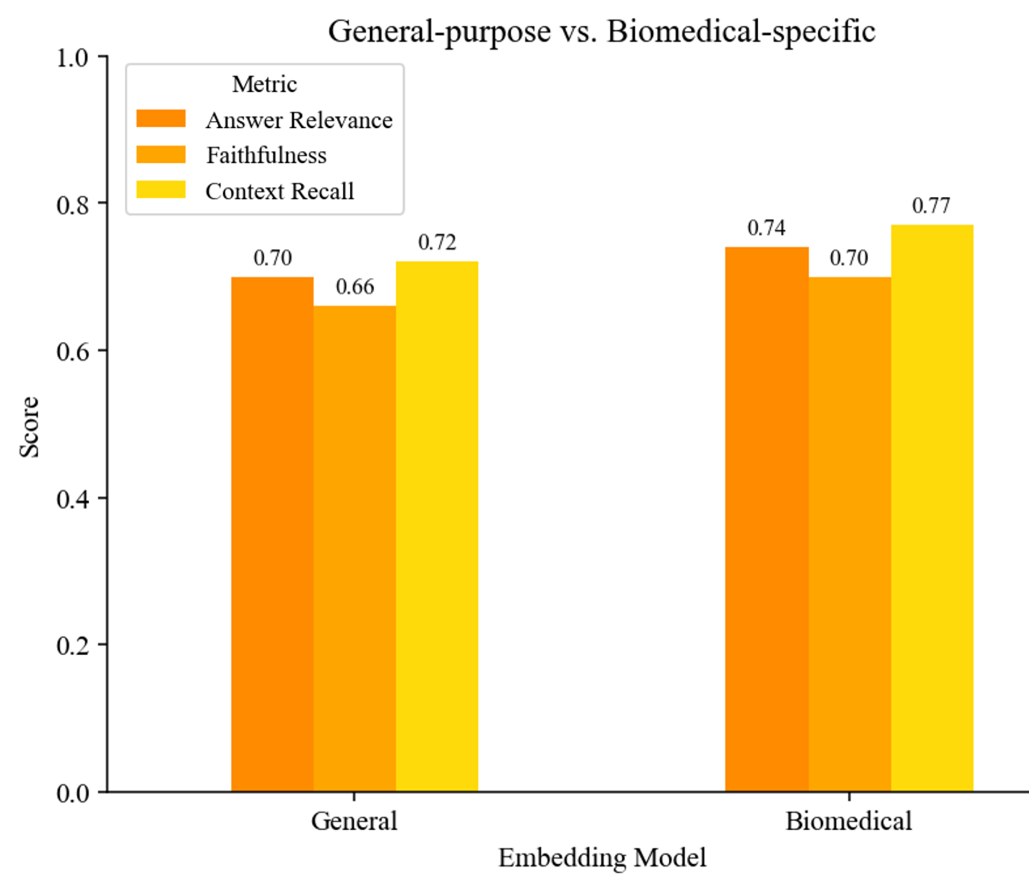}
    \captionsetup{justification=centering}
    \caption{Results of the embedding model comparison (General vs. Biomedical), with recursive 512-token chunking and hybrid search retrieval (Experiments 8 and 10).}
    \label{fig:embedding}
\end{figure}

Experiment 10 (Figure \ref{fig:embedding}) demonstrated that replacing the general-purpose model with a biomedical-specific embedding model further improved all evaluated metrics. Comparing to the best-performing configuration from the previous experiments (recursive 512-token chunks with hybrid search retrieval), the biomedical embedding model boosted answer relevance (0.74 vs. 0.70), faithfulness (0.70 vs. 0.66), and context recall (0.77 vs. 0.72).

This result highlights the critical importance of domain-adapted embeddings in clinical applications. General-purpose embeddings, while efficient, often lack the granularity to capture subtle distinctions in biomedical terminology (e.g., reagent abbreviations, histological stain variations). By contrast, MedEmbed demonstrated a superior capacity to align user queries with the precise procedural contexts needed in AP labs. These findings strongly support ongoing efforts to develop and fine-tune domain-specific embedding models for healthcare, as they can directly improve retrieval accuracy, reduce hallucinations, and enhance trustworthiness of AI assistants.

\subsection{Top-$k$ evaluation results}
As shown in Table \ref{tab:top_k}, the top-$k$ retrieval experiment highlighted a clear trade-off between precision and recall. As the number of retrieved chunks ($k$) increased, the system captured a higher proportion of relevant information (improving recall) at the expense of introducing more irrelevant or redundant chunks (reducing precision).

\begin{table}[!ht]
    \centering
    \caption{Top-$k$ evaluation results.}
    \label{tab:top_k}
    \begin{tabular}{@{}cccc@{}}
        \toprule
        \textbf{$k$} & \textbf{Precision@k} & \textbf{Recall@k} & \textbf{F1-Score@k} \\
        \midrule
        $k=1$ & \textbf{0.45} & 0.59 & \textbf{0.50} \\
        $k=2$ & 0.32 & 0.69 & 0.41 \\
        $k=4$ & 0.19 & 0.76 & 0.29 \\
        $k=8$ & 0.12 & \textbf{0.83} & 0.19 \\
        \bottomrule
    \end{tabular}
\end{table}

In our evaluation, as demonstrated by the highest F1 score (0.50), retrieving a single top-ranked chunk ($k=1$) provided the best overall balance. Increasing $k$ beyond 1 tended to add context that was often unnecessary, which lowered the precision of retrieval without significantly improving answer quality. 

This result has two key implications. First, it demonstrates that AP protocols are highly modular, with most questions answerable from a single protocol excerpt rather than requiring cross-document aggregation. Second, it suggests that retrieving more than a few chunks may dilute relevance and reduce efficiency, particularly under hardware or latency constraints. However, in scenarios where tasks require synthesizing knowledge across multiple chunks (e.g., multi-step staining protocols), higher values of $k$ could still prove beneficial.

\subsection{Comparison with Previous Works}
To contextualize our findings, we compare our results with recent RAG implementations in healthcare and technical documentation domains. Our finding that recursive chunking (512 tokens) outperformed semantic chunking contrasts with results reported in some general-domain studies \cite{Jiang2024}, where semantic chunking showed advantages for narrative text. However, this aligns with findings from technical documentation retrieval studies that emphasize the importance of consistent chunk sizes for procedural content \cite{Finardi2024}. The modular, step-wise structure of AP protocols, where each section contains discrete procedural instructions, appears better suited to fixed-length segmentation than semantic boundary detection.

Concerning the retrieval method effectiveness, the superior performance of hybrid search is consistent with recent healthcare RAG implementations \cite{gargari2025, amugongo2025}, which have similarly demonstrated that combining dense and sparse retrieval improves recall without significantly compromising precision. Our hybrid search configuration (30\% semantic, 70\% keyword weighting) differs from the 50-50 balance commonly reported in literature. This adjustment was necessary to accommodate the high density of technical terminology in AP protocols, where exact term matching is frequently essential.

The substantial performance gains from biomedical-specific embeddings (MedEmbed) validate findings from BioBERT \cite{Lee2020} and recent medical information retrieval studies \cite{gargari2025}. Specifically, our improvement in answer relevance (0.74 vs. 0.70) is comparable to the 4-8\% improvements reported in clinical QA tasks using domain-adapted models. This highlights a broader pattern: domain specialization in embedding models produces proportionally greater benefits in highly technical domains with specialized vocabularies.

Our finding that $k=1$ maximizes F1-score (0.50) diverges from studies where larger $k$ values ($k=3$-$5$) typically perform better \cite{Lewis2021}. This distinction reflects fundamental differences in query characteristics: whereas clinical decision support often requires synthesizing information across multiple sources (e.g., patient history, treatment guidelines, drug interactions), laboratory protocol queries typically seek single, authoritative procedural specifications. This finding suggests that retrieval optimization strategies must be tailored to task-specific information needs rather than applying uniform configurations across domains.

Finally, our RAGAS scores (faithfulness: 0.70, answer relevance: 0.74, context recall: 0.77) are comparable to or exceed benchmarks reported in recent medical RAG evaluations \cite{amugongo2025}, which typically achieve faithfulness scores between 0.65-0.72 and answer relevance between 0.68-0.75 on clinical QA tasks. However, direct comparison is complicated by differences in corpus characteristics, query complexity, and ground-truth construction methods. Our use of protocol-derived QA pairs may represent a more controlled evaluation setting compared to real-world clinical queries, potentially contributing to our relatively strong performance.

\subsection{Main Findings}
All in all, the results presented in this section (summarised in Table \ref{tab:all_results}) yield several fundamental insights into building RAG systems for AP.
With respect to the chunking strategy, recursive chunking with moderate token lengths (512 tokens) consistently outperformed semantic approaches across all evaluation metrics. This finding contradicts assumptions that semantic boundaries are inherently superior for technical documents. The poor performance of semantic chunking resulted from highly variable chunk sizes that diluted relevant information concentration. For AP protocols, which follow standardised structural patterns, length-based chunking with appropriate overlap proved more effective. 

Hybrid search combining semantic and keyword matching achieved the best overall performance, particularly for context recall. However, naïve embedding-based retrieval maintained slightly higher faithfulness scores, revealing an inherent trade-off between coverage and grounding in RAG systems. The reranking approach failed to improve upon baseline performance, suggesting that threshold-based filtering may be suboptimal for highly specific technical queries.

The biomedical-specific embedding model (MedEmbed) provided substantial improvements across all metrics compared to general-purpose embeddings. The improvement in context recall and answer relevance demonstrates the critical importance of domain adaptation in clinical AI systems. These gains likely reflect better handling of specialised terminology, abbreviations, and contextual relationships specific to laboratory procedures.

Finally, the superior performance of single-chunk retrieval reflects the modular, self-contained nature of AP protocols. This finding has significant practical implications for deployment, as it reduces computational overhead while maintaining accuracy. This is especially critical in resource-constrained laboratory environments, where minimizing latency is as important as ensuring accuracy.

The broader significance of these findings is twofold. Methodologically, they provide empirical evidence that design choices in chunking, retrieval, and embeddings substantially shape the reliability of RAG systems in healthcare. Practically, they show that with careful optimisation, RAG assistants can transform static protocol documentation into dynamic, queryable knowledge, directly supporting technicians in reducing errors and maintaining diagnostic reliability.

More broadly, this work illustrates how RAG assistants can serve as trusted digital companions within laboratory medicine, allowing for a real-time procedural support. By enabling rapid, reliable access to the latest validated protocols, such systems have the potential to contribute to more consistent and accurate diagnoses, with downstream benefits for patient outcomes.

\begin{table}[h!]
    \centering
    \caption{Overall experiment results.}
    \label{tab:all_results}
    \begin{tabular}{clccc}
        \toprule
        \textbf{Exp.} & \textbf{Description} & \textbf{Answer Relevance} & \textbf{Faithfulness} & \textbf{Context Recall} \\
        \midrule
        1  & Naïve – 256 chunks (baseline)         & 0.68 & 0.65 & 0.52 \\
        2  & Naïve – 512 chunks                    & 0.64 & 0.62 & 0.69 \\
        3  & Naïve semantic chunks                 & 0.52 & 0.48 & 0.33 \\
        4  & Reranking – 256 chunks                & 0.69 & 0.64 & 0.58 \\
        5  & Reranking – 512 chunks                & 0.63 & 0.63 & 0.68 \\
        6  & Reranking semantic chunks             & 0.51 & 0.31 & 0.23 \\
        7  & Hybrid search – 256 chunks            & 0.69 & 0.65 & 0.65 \\
        8  & Hybrid search – 512 chunks            & 0.70 & 0.66 & 0.72 \\
        9  & Hybrid search – semantic chunks       & 0.67 & 0.30 & 0.31 \\
        10 & Hybrid search – 512 biomedical chunks & \textbf{0.74} & \textbf{0.70} & \textbf{0.77} \\
        \bottomrule
    \end{tabular}
\end{table}

\section{Conclusions and Future Work}
This study demonstrates the viability and effectiveness of Retrieval-Augmented Generation systems for supporting clinical laboratory workflows in Anatomical Pathology. Through systematic evaluation of design choices spanning chunking strategies, retrieval methods, and embedding models, we identified an optimal configuration that combines recursive 512-token chunking, hybrid semantic-keyword retrieval, and biomedical-specific embeddings (MedEmbed). This configuration achieved clinically relevant performance levels: faithfulness (0.70), answer relevance (0.74), and context recall (0.77).
Our experimental findings provide three key insights for RAG deployment in technical healthcare domains. First, simple length-based chunking with moderate token counts ($\sim$512) consistently outperforms semantic approaches for procedural documentation with standardized structure. Second, hybrid retrieval strategies that balance semantic understanding (30\%) with keyword matching (70\%) optimize recall without sacrificing precision, particularly critical when queries use terminology variants. Third, domain-adapted embeddings substantially improve performance over general-purpose models, validating the necessity of specialized AI tools for clinical applications. The superior performance of single-chunk retrieval ($k=1$) further demonstrates that AP protocols' modular structure enables accurate responses from precisely targeted context, reducing computational overhead while maintaining accuracy, an important consideration for resource-constrained laboratory environments.
Several limitations must be acknowledged when interpreting these results. First, our evaluation dataset comprised synthetically generated question-answer pairs derived directly from protocol content, which may not fully capture the complexity, ambiguity, and contextual variability of real-world technician queries, particularly those involving troubleshooting, exceptions, or integration across multiple protocols. Second, our corpus originated from a single Portuguese healthcare institution, potentially limiting generalizability across different laboratory settings, equipment configurations, or international practice standards. Third, hardware constraints limited our exploration to smaller models (Llama 3.1 8B), and evaluation was conducted exclusively on Portuguese-language documents, raising questions about transferability to other languages and larger model architectures.
Future work should prioritize multi-institutional validation studies involving real technician queries and longitudinal deployment in active laboratory settings to assess practical impact on workflow efficiency and error rates. Extension to other clinical laboratory specialties (clinical chemistry, microbiology, molecular diagnostics) would establish the broader applicability of our design principles. Additionally, investigating multilingual capabilities, integration with voice-based query interfaces for hands-free operation, and incorporation of real-time protocol update mechanisms represent promising directions for enhancing practical utility and ensuring that RAG assistants remain synchronized with evolving laboratory practices.


\bibliographystyle{unsrtnat}
\bibliography{references}

\subsection*{Ethics statement}
All procedures were performed in compliance with relevant laws and institutional guidelines and have been approved by the institutional board on the 22nd of May 2025. Ethical approval was not necessary considering that no human subjects and data were used in this study.

\subsection*{Data availability}
The corpus datasets that support the findings of this study are openly available in HuggingFace at \url{https://huggingface.co/datasets/diogofmp/AP_Lab_Protocols}. To replicate the study, the code used is available on GitHub at \url{https://github.com/diogo-pires-github/RAG_for_biomedical_protocols}.



\subsection*{Competing interests}
The authors declare no competing interests.

\end{document}